 \documentclass[grl]{agutex}

 \usepackage[dvips]{graphicx}
 \setkeys{Gin}{draft=false}
\usepackage{url}

\authorrunninghead{GACESA ET AL.}

\titlerunninghead{Non-thermal escape of H$_2$ from Mars}

\begin{document}


\title{Non-thermal escape of molecular hydrogen from Mars}
%
%




\authors{M. Gacesa\altaffilmark{1,2}, 
P. Zhang\altaffilmark{1}, and V. Kharchenko\altaffilmark{1,2}}

\altaffiltext{1}{Institute for Theoretical Atomic and Molecular Physics, 
Harvard-Smithsonian Center for Astrophysics, Cambridge, MA, USA.}

\altaffiltext{2}{Department of Physics, University of Connecticut,
Storrs, CT, USA.}



\begin{abstract}
We present a detailed theoretical analysis of a non-thermal escape of molecular hydrogen from Mars induced by collisions with hot atomic oxygen from martian corona. To accurately describe the energy transfer in O + H$_2(v,j)$ collisions, we performed extensive quantum-mechanical calculations of state-to-state elastic, inelastic, and reactive cross sections. The escape flux of H$_2$ molecules was evaluated using a simplified 1D column model of the martian atmosphere with realistic densities of atmospheric gases and hot oxygen production rates for the low solar activity conditions. 
An average density of the non-thermal escape flux of H$_2$ of $1.9\times10^5$ cm$^{-2}$s$^{-1}$ was obtained considering energetic O atoms produced in dissociative recombinations of O$_{2}^{+}$ ions. Predicted rovibrational distribution of the escaping H$_2$ was found to contain a significant fraction of higher rotational states.
While the non-thermal escape rate was found to be lower than Jeans flux for H$_2$ molecules, the non-thermal escape rates of HD and D$_2$ are significantly higher than their respective Jeans rates. 
The accurate values of non-thermal escape fluxes of different molecular isotopes of H$_2$ may be important in analyses of evolution of the martian atmosphere.
The described molecular ejection mechanism is general and expected to contribute to atmospheric escape of H$_2$ and other light molecules from planets, satellites, and exoplanetary bodies.
\end{abstract}


\begin{article}


\section{Introduction}
The interaction of the martian atmosphere with the solar radiation and interplanetary plasma results in its evaporation due to thermal (Jeans) escape and a number of non-thermal mechanisms. The absence of the intrinsic magnetic field on Mars and its low gravitational potential make the martian atmosphere particularly susceptible to erosion \citep{1998Sci...279.1676A}. The current low atmospheric pressure can be explained well by extrapolating the escape rates to a geological time frame while accounting for a change of the solar activity in the past \citep{2004P&SS...52.1039C}. 
The present day degradation of the martian atmosphere occurs mainly via non-thermal escape processes induced by ion charge-exhange, sputtering and ionospheric outflows driven by solar wind \citep{2004P&SS...52.1039C,2008SSRv..139..355J}. 
The dissociative recombination (DR) of O$_{2}^{+}$ is a major source of hot O atoms in the upper atmosphere of Mars, responsible for the escape of oxygen and formation of martian hot corona \citep{1988Icar...76..135I,1993GeoRL..20.1747F,2005SoSyR..39...22K}.

Nascent hot O atoms collide with thermal constituents of the martian atmosphere and eject them from the planetary gravitational field, if a sufficient kinetic energy transfer occurs. Suprathermal neutral oxygen was shown to be important for analyses of Mars' corona and the non-thermal escape of neutral atoms \citep{2005SoSyR..39...22K,2006SoSyR..40..384K,2009Icar..204..527F}. Recent calculations of the non-thermal He escape from Mars carried out with accurate energy transfer parameters predicted a significant He escape flux induced by hot O atoms \citep{2011GeoRL..3802203B}. 

In this Letter we explore collisional ejection of molecules from the martian atmosphere. Specifically, we report the results of a quantum-mechanical study of the energy transfer from the hot O to H$_2$ molecules and their subsequent escape. Significant computational difficulties arise from the fact that molecular internal rotational and vibrational degrees of freedom can be excited in collisions. In addition, the reactive pathway leading to the production of OH molecules is energetically permitted. To account for the increased complexity, we have computed the cross sections for O($^3$P) + H$_2$ reactive collision using fully quantum-mechanical approach. Kinetic theory was used to calculate the rate of energy transfer, as well as distributions of excited rotational and vibrational (RV) states of the recoiled H$_2$ molecules. The total escape flux of H$_2$ from Mars and RV distributions of escaping molecules have been evaluated for the low solar activity conditions. Also, we have estimated the non-thermal escape fluxes of HD and D$_2$ and compared them to the corresponding Jeans escape rates.
Finally, the dependence of the molecular ejection fluxes on the gravitational escape threshold is analyzed for conditions present on other planets, satellites, and exoplanets.

\section{Cross Sections and Energy Transfer}

The DR of O$_{2}^{+}$ with electrons proceeds via five possible dissociation pathways, producing O($^3$P), O($^1$D), and O($^1$S) \citep{1988P&SS...36...47G,2009Icar..204..527F}. 
Energetic metastable O($^1$D) atoms decay via spontaneous emission and quenching in collisions with atmospheric gases into O($^3$P) atoms \citep{2005JGRA..11012305K}. The cross sections for O($^3$P) and O($^1$D) colliding with He were found to be very similar \citep{2011GeoRL..3802203B}. 
For simplicity, we assumed a similar behavior for O($^3$P) + H$_2$ and O($^1$D) + H$_2$ elastic collisions.

To describe the collision ejection of H$_2$ molecules, we have calculated elastic and inelastic cross sections for the center-of-mass (CM) collision energies from 0.01 to 4.5 eV. The quantum scattering code ABC \citep{2000CoPhC.133..128S}, that can treat elastic, inelastic, and open reactive channels of the OH production, $\mathrm{O}(^3P) + \mathrm{H}_2(v,j) \rightarrow \mathrm{OH}(v'',j'') + \mathrm{H}$, where $(v,j)$ and $(v'',j'')$ indicate initial and final RV levels of H$_2$ and OH, was used to solve the time-independent coupled-channel Schr\"odinger equation in Delves hyperspherical coordinates. 
In addition, for high collision energies, the elastic and inelastic cross sections were calculated using the MOLSCAT \citep{MOLSCAT} code to ensure the convergence of the nonreactive channels\footnote{A detailed description of the scattering calculations and resulting cross sections for the two lowest potential energy surfaces will be published elsewhere.}. Extensive numerical convergence tests were carried out for the both codes.

\begin{figure}[htbp]
\noindent \includegraphics[width=20pc]{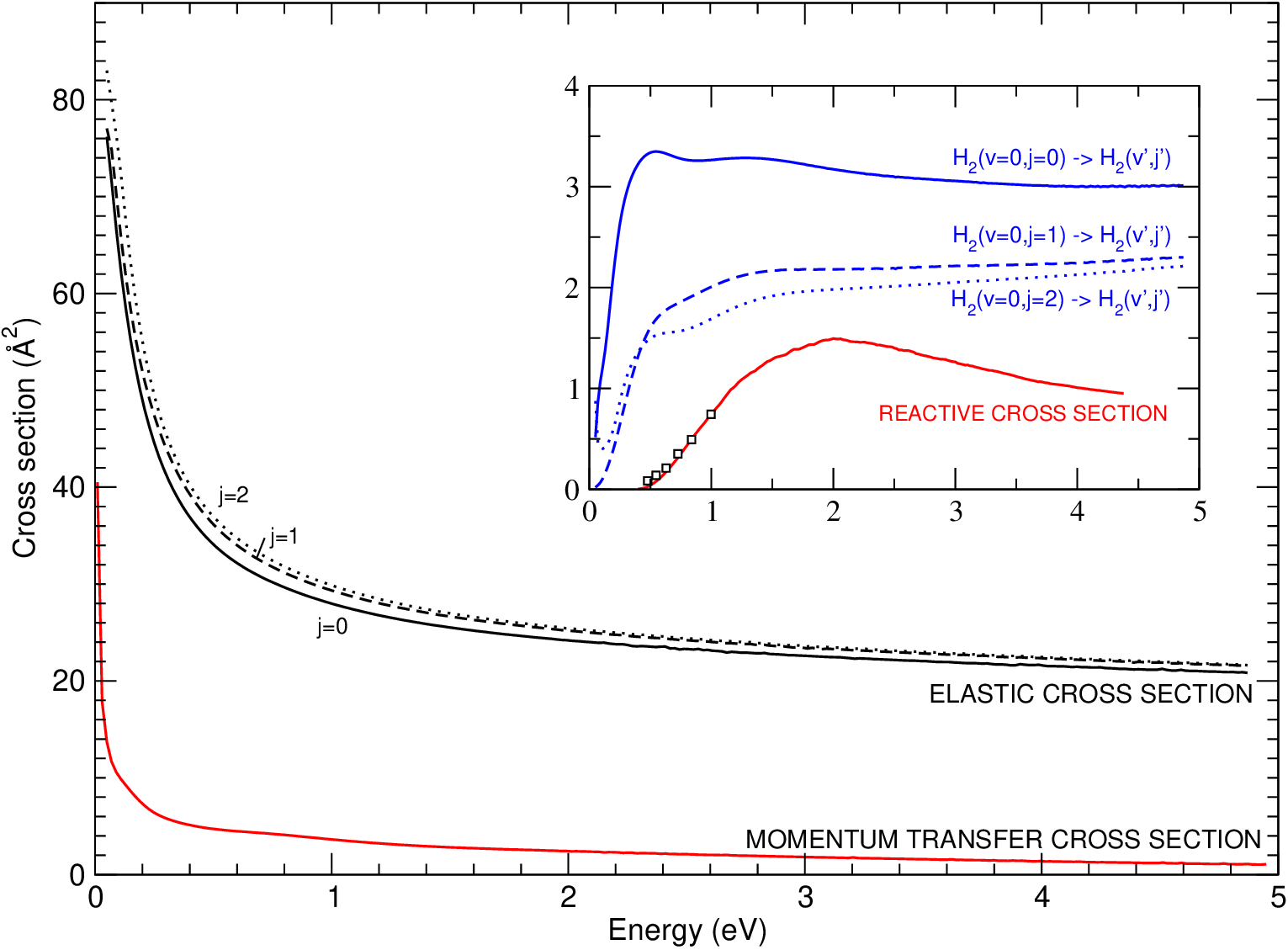}
\caption{
Elastic and momentum transfer cross sections for H$_2(v=0,j=0,1,2)$ + O collisions. The momentum transfer cross section shown is thermally averaged over the first three rotational states.
\textit{Inset:} Total inelastic cross sections $\sigma_{vj}^{\mathrm{inel}}(E) = \sum_{v',j'}\sigma_{vj,v'j'}(E)$ for H$_{2}(v=0,j=0,1,2) + \mathrm{O} \rightarrow \mathrm{H}_{2}(v'=0,j')$ + O, and $j \neq j'$.
The reactive cross section for H$_{2}(v=0,j=0) + \mathrm{O} \rightarrow \mathrm{OH} + \mathrm{H}$ is also shown and compared with the experimental results (black squares) \citep{2003JChPh.118.1585G}.}
\label{fig1}
\end{figure}

The O($^3P$)+H$_2(v,j)$ interaction was described using two lowest potential energy surfaces, Rogers' LEPS $^3A''$ \citep{Rogers_Kupperman_PES_2000} and Brand\~ao's BMS1 $^3A'$ \citep{2004JChPh.121.8861B}. 
Partial cross sections for initial and final rotational levels $j$ and $j'$ were constructed as a statistically weighted sum of the independently calculated cross sections for the two potential surfaces, where both $^3A''$ and $^3A'$ contribute a weight factor of $1/3$ \citep{2004JChPh.121.6346B}. Elastic, inelastic, and momentum transfer partial cross sections for oxygen colliding with the hydrogen molecule in three energetically lowest rotational states are given in Figure \ref{fig1}.
We compared our reactive cross sections for OH production to the previously published results \citep{2003JChPh.119..195B,2004JChPh.121.6346B,2004JChPh.120.4316B,2010ChJCP..23..149W} 
and found them to be in close agreement within the available energy range. 

To determine the energy transfer rate from the suprathermal oxygen to atmospheric H$_2$ and find its escape rate, we used kinetic theory with a quantum description of internal molecular structure and realistic anisotropic cross sections. Since the reactive cross sections are an order of magnitude smaller than the elastic cross sections (Figure \ref{fig1}), and the more massive OH molecule has a considerably higher escape threshold than H$_2$, we neglected it in this study. However, note that a small fraction of produced OH molecules may be sufficiently energetic to escape.
The transferred energy from the energetic projectile O to the frozen target H$_2$ in the laboratory frame (LF) can be expressed as \citep{1982itam.book.....J}
\begin{equation}
  T_{v',j'} = \frac{m_\mathrm{O} \, m_{\mathrm{H}_2}} {(m_\mathrm{O} + m_{\mathrm{H}_2})^2} 
      \left( 1 + \gamma_{v',j'} - 2\sqrt{\gamma_{v',j'}} \cos \theta \right) E,
\label{eq1}
\end{equation}
where 
$m_\mathrm{O}$ and $m_\mathrm{{H_2}}$ are masses of O and H$_2$, respectively, $E$ is the collision energy in the LF, $\gamma_{v',j'} = \epsilon_{v'j'} / \epsilon$, is the ratio of $\epsilon_{v'j'}$ and $\epsilon$, the CM translational kinetic energies after and before the collision, respectively, and $\theta$ is the scattering angle in the CM frame. 
The energies $\epsilon_{v'j'}$ were calculated quantum mechanically for the two triplet potential surfaces. Eq. (\ref{eq1}) takes into account that the energy transferred to H$_2$ molecules is spent on increasing their translational kinetic energy and exciting their internal RV degrees of freedom.

The fraction of energized H$_2$ molecules capable of escaping can be calculated as 
\begin{equation}
  \Gamma_{vj}^{v'j'}(E) = \frac{\int_{\theta_{\mathrm{min}}}^{\pi} Q_{vj,v'j'}(\theta) 
                 \left(1-\cos \theta \right) \sin \theta d \theta }
             {\int_{0}^{\pi} Q_{vj,v'j'}(\theta) \left(1-\cos \theta \right) \sin \theta d \theta } ~,
  \label{eq2}
\end{equation}
where $Q_{vj,v'j'}(\theta)$ is the differential cross section for scattering of H$_2$ in the initial $(v,j)$ into the final $(v',j')$ state. 
The critical angle $\theta_{\mathrm{min}}$ was determined from the condition that the translational part of the transferred energy $T_{v',j'}$ is equal to the minimum energy required for H$_2$ to escape from Mars, $E_{\mathrm{esc}} = 0.26$ eV. An alternative description of the escape process could be constructed by performing Monte Carlo simulations with accurate quantum cross sections for angular distributions of the recoiled H$_2$ molecules.

Momentum transfer cross sections $\sigma^{\mathrm{mt}}_{v j, v' j'}$ for inelastic collisions were calculated using \citep{1978JChPh..68.1585P}
\begin{equation}
  \sigma^{\mathrm{mt}}_{v j, v' j'} = 2 \pi \int_0^{\pi} \mathrm{d} \theta \sin \theta 
       \left( 1-\sqrt{ 1 - \gamma_{v',j'}} \cos \theta \right) Q_{vj,v'j'}(\theta) .
\label{eq3}
\end{equation}


\section{Flux and Distribution of Escaping H$_2$}

\begin{figure}[htp]
\noindent\includegraphics[width=20pc]{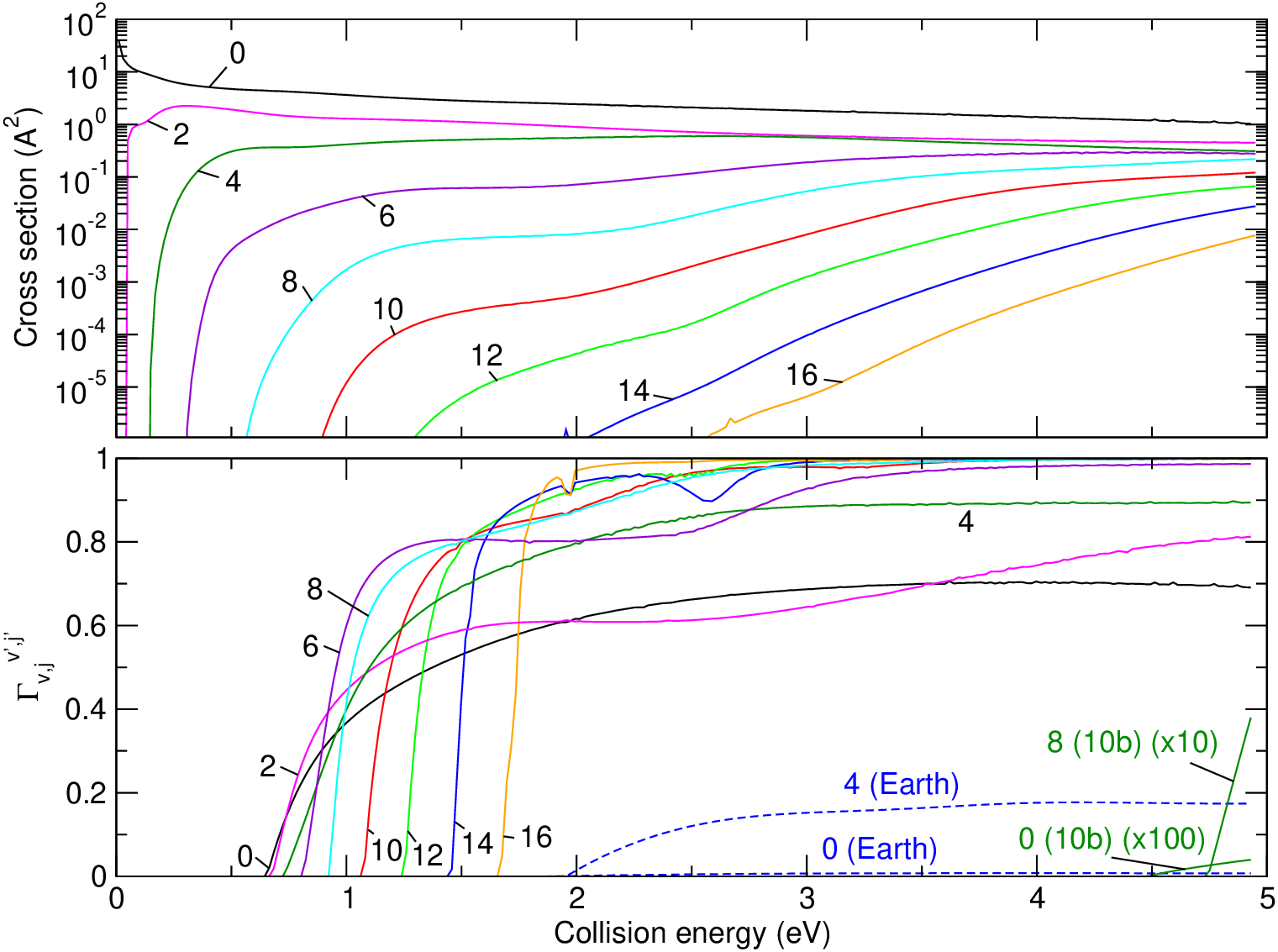}
\caption{\textit{Top:} State-to-state momentum transfer cross sections for the H$_{2}(v',j')$. Curves for the final rotational levels, $j'=0,2,\ldots,16$ (labeled on graph), collisionally excited from the ground state of H$_{2}$, are shown.
\textit{Bottom:} Fraction $\Gamma_{v j}^{v' j'}$ of the recoiled H$_{2}(v'=0,j')$ with energies greater than the escape energy for Mars. The escaping fractions for Earth for $j'=0,4$ (dashed blue) and Kepler-10b super-earth for $j'=0,8$ (light green) are given for comparison.}
\label{fig2}
\end{figure}

Jeans escape and collisions with hot oxygen are the major mechanisms that contribute to the escape of neutral H$_2$ molecules and their isotopes from the martian atmosphere.
Both processes are strongly dependent on the temperature and density of upper layers of the martian atmosphere. The temperature of the exosphere (above the altitude of about 180 km), $T_{\mathrm{exo}}$, is approximatelly constant and estimated to be between 240 and 280 K, depending on the solar activity and gas density profiles \citep{2010Icar..207..638K,2009Icar..204..527F,2003JGRA..108.1223F}. To obtain a conservative estimate of the non-thermal flux of escaping H$_2$, we considered the $T_{\mathrm{exo}} = 240$ K, corresponding to the low solar activity.
Furthermore, we assumed a thermal distribution of the initial rotational states of H$_2$, where more than 95 \% of the total population is distributed between its first three rotational states, $j=0,1,2$, with corresponding population fractions equal to 0.31, 0.46, and 0.19, respectively. 
Using Eqs. (\ref{eq2},\ref{eq3}) we have calculated the values of the thermally averaged momentum transfer cross sections and the fractions $\Gamma_{vj}^{v'j'}(E)$ of rovibrationally excited H$_{2}$, sufficiently energetic to escape from Mars (Figure \ref{fig2}). 
The fraction $\Gamma_{vj}^{v'j'}(E)$ becomes significant at collision energies greater than 0.7 eV for H$_2(j'=0,2)$. Although higher rotational states require increasingly larger projectile energies, \textit{e.g.} H$_2(j'=16)$ can escape for $E>1.65 \; \mathrm{eV}$, their fraction in the RV distribution of the escaping molecules also becomes larger. Note that the initial population of higher vibrational levels of H$_2$ at the exobase is negligible.

Since $\Gamma_{vj}^{v'j'}(E)$ depends only on the energy transfer efficiency and the escape energy threshold, it can be easily generalized to different astronomical objects. We illustrate this for two hypothetic planets, the first corresponding in size and mass to Earth and the second to the extrasolar super-earth Kepler-10b (3.3 Earth masses) \citep{0004-637X-729-1-27}. Note that only very energetic H$_2$, mostly in higher excited rotational levels, is able to escape (Figure \ref{fig2}).

A simple estimate of the total escape flux of H$_{2}(v',j')$ can be obtained from the exobase approximation \citep{2003JGRA..108.1223F,2004P&SS...52.1039C,2010Icar..207..638K}, using the density of exospheric H$_2$, and fractions $\Gamma_{vj}^{v'j'}(E)$ calculated above. However, such an approach neglects the hot O production below the exobase and loss of the upward flux in atmospheric collisions, resulting in a large uncertainty in the computed flux.

We constructed a more realistic 1D model of escape, analogous to the one used to describe the escape of He atoms \citep{2011GeoRL..3802203B}. In our model the explicit consideration of energy transfer collisions is combined with the altitude-dependent rate of production of hot O atoms, $f(E,h)$ via DR channels \citep{1988P&SS...36...47G,2005JChPh.122w4311P,2011GeoRL..3802203B}. In addition, we estimated the extinction of fluxes of suprathermal O and H$_2$ due to collisions with thermal atmospheric gases. All calculations were performed for low solar activity. We used the rate of production of hot O below 400 km by \citet{2009Icar..204..527F} and smoothly interpolated it to the rate given by \citet{1996JGR...10115765K} at higher altitudes.

The volume production rate of escaping hot H$_2(v',j')$ can be expressed as
\begin{eqnarray}
  P_{v' j'}(h) & = & \frac{1}{2} \int_{0}^{\infty} \mathrm{d}E \, T_{\mathrm{H}_2}(h,E) 
      n_{\mathrm{H}_2}(h) \Gamma_{vj}^{v'j'}(E) \sigma_{vj,v'j'}^{\mathrm{mt}}(E) \nonumber  \\
             &   & \times \int_{h_{2}^{\mathrm{min}}}^{h} \mathrm{d}h_2 f(E,h_2) T_\mathrm{O}(h_2,h,E) ,
  \label{eq:P}
\end{eqnarray}
with the transparency factors $T_{\mathrm{H}_2}$ and $T_\mathrm{O}$ defined as
\begin{eqnarray}
  T_{\mathrm{H_2}}(h,E) & = & \exp \left[-\int_{h}^{h_\mathrm{max}} \mathrm{d}h' \sum_i
                 \sigma_{\mathrm{H_2},i}^{\mathrm{mt}}(E) n_{i}(h') \right] \nonumber \\
  T_\mathrm{O}(h_2,h,E) & = & \exp \left[ -\int_{h_2}^{h} \mathrm{d}h' \sum_i 
                      \sigma_{\mathrm{O},i}^{\mathrm{mt}}(E) n_{i}(h') \right] .
\end{eqnarray}

\begin{figure}[htbp]
\noindent\includegraphics[width=20pc]{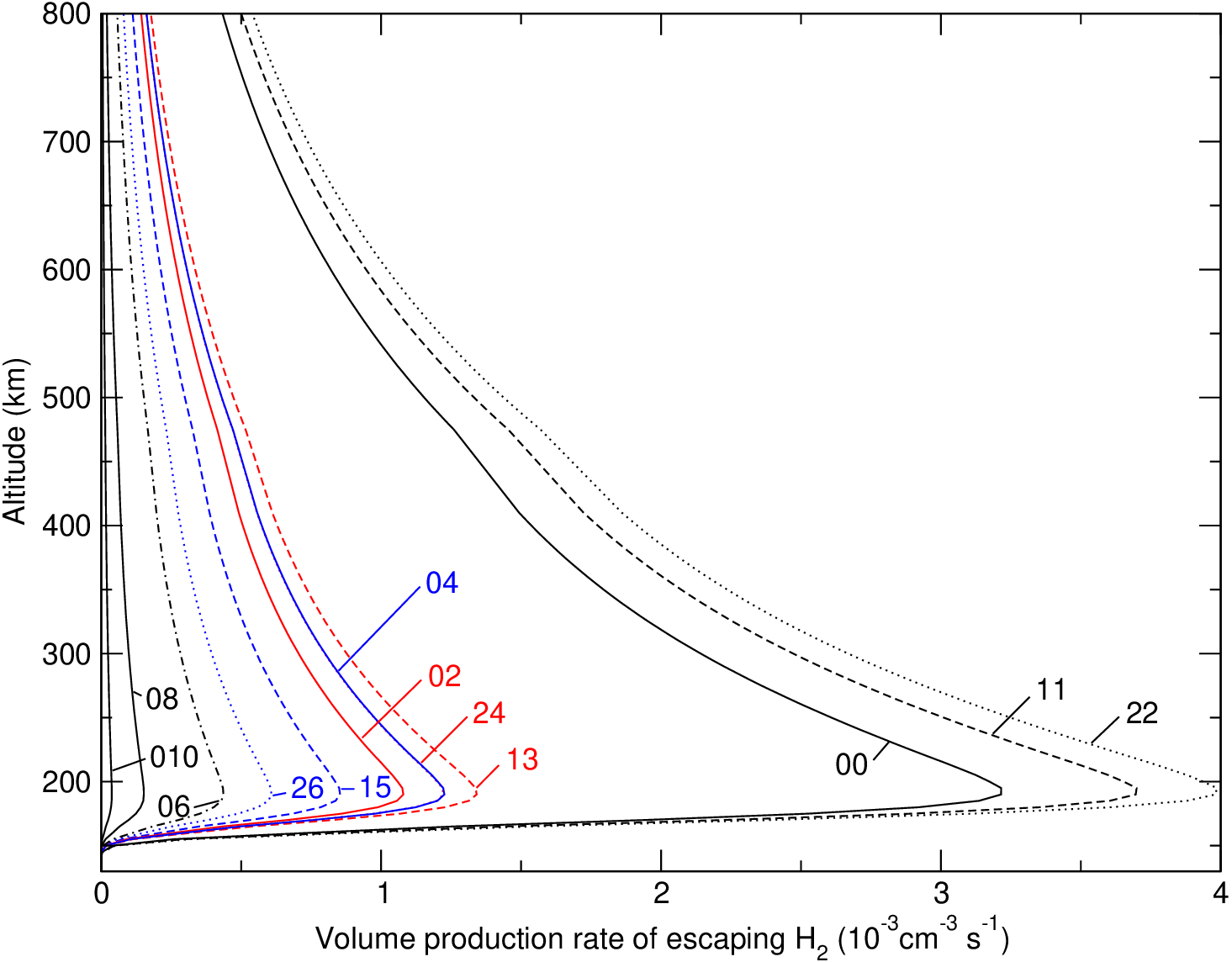}
\caption{Altitude profile of the volume production rate $P_{v',j'}(h)$ of the non-thermal flux of H$_2$ molecules escaping from Mars. The most significant rates with respect to the initial and final rotational states, $j=0$ (solid), $j=1$ (dashed) and $j=2$ (dotted), and $j'=0-10$, are shown. The curves are denoted as $jj'$.}
\label{fig3}
\end{figure}

The transparency factor $T_{\mathrm{H}_2}$ is equal to the escape probability of hot H$_2(v',j')$ produced in collisions with the incident hot O of energy $E$ at the altitude $h_2$. 
The second transparency factor $T_\mathrm{O}$ is defined as the probability that the hot O atoms, produced at the altitude $h_2$, reach the altitude $h$ without the energy loss in collisions with other atmospheric constituents. 
The quantity $n_{\mathrm{H}_2}(h) \, \Gamma_{vj}^{v'j'}(E) \, \sigma_{vj,v'j'}^{\mathrm{mt}}(E)$ is the inverse mean free path for O+H$_2(v,j)$ collisions, resulting in the energy transfer greater than the H$_2$ escape threshold. 
The prefactor $1/2$ indicates that, in our simplified 1D model, approximately half of the nascent energetic atoms and recoiled H$_2$ molecules are scattered towards the planet and cannot escape regardless of the energy transferred.
Summations of the flux loss of H$_2$ and O in collisions with the $i$-th atmospheric gas of density $n_i(h)$ and momentum transfer cross section $\sigma_{\mathrm{H_2},i}^{\mathrm{mt}}(E)$ and $\sigma_{\mathrm{O},i}^{\mathrm{mt}}(E)$, respectively, included major constituents of the martian upper atmosphere: CO$_2$, CO, N$_2$, O$_2$, H$_2$, H, Ar, and He. 
The momentum transfer cross sections for H-H$_2$ \citep{1999JPhB...32.2415K}, Ar-H$_2$ \citep{2005JChPh.122b4304U}, H$_2$-H$_2$ \citep{1990JPCRD..19..653P} were used from the literature. Since no data were available in the required energy range, we used approximate mass-scaled cross sections for He-H$_2$ (from Ar-H$_2$), N$_2$-H$_2$, O$_2$-H$_2$, CO-H$_2$ (from O-H$_2$), and CO$_2$-H$_2$ (from O-N$_2$ \citep{1998JGR...10323393B}).

Using Eq. (\ref{eq:P}) we have calculated volume production rate of the escaping H$_2(v'=0,j')$ molecules induced in H$_2(v=0,j=0-2)$ + O collisions for a range of altitudes from $h_{\mathrm{min}}=130$ km to $h_{\mathrm{max}}=800$ km (Figure \ref{fig3}). Note that, by symmetry arguments, for a homonuclear H$_2$ only $\Delta j=0,2,4...$ transitions are allowed \citep{1986qmv2.book.....C}.
The resulting escape rates of H$_2$ molecules are the largest for the elastic collisions, followed by the three times smaller rates for the first two excited rotational states, $j'=2$ and $j'=4$. The rates remain significant for the final rotational levels up to $j'=10$. 
\begin{figure}[htbp]
\noindent\includegraphics[width=20pc]{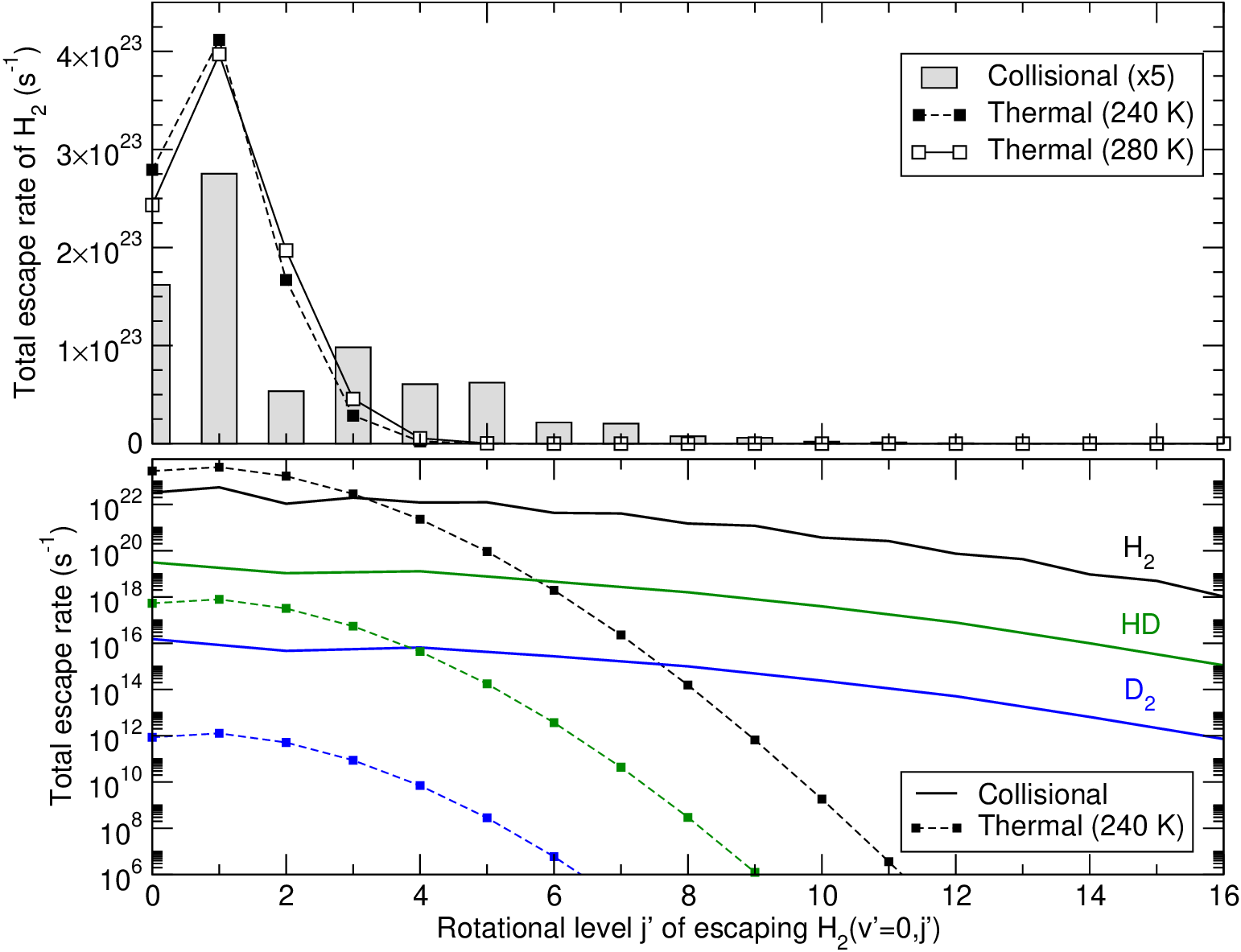}
\caption{\textit{Top:} Collisional and thermal total escape rate of H$_2(v'=0,j')$ for the first 16 rotational states $j'$. \textit{Bottom:} The same as above for H$_2$, HD, and D$_2$.}
\label{fig4}
\end{figure}
The altitude profile of the production rate of H$_2$ capable of escaping is similar to the production rate profile of He \citep{2011GeoRL..3802203B}. This was expected, since the escape of both species is driven by collisions with the nascent fast O atoms, produced mostly below 150 km for the considered atmospheric and solar conditions. The calculated altitude profile can be used to compute the non-thermal escape flux of H$_2$ molecules and estimate the accuracy of the exobase approximation.
We calculated $\phi_{j'}$, the non-thermal flux for H$_2(v'=0,j')$ molecules as 
\begin{equation}
  \phi_{j'} = \int_{h_{\mathrm{min}}}^{h_{\mathrm{max}}} P_{v'j'}(h) \mathrm{d}h .
\end{equation}
Total collisional and thermal fluxes were calculated as sums over all rotational levels and found to be $1.9 \times 10^5$ cm$^{-2}$ s$^{-1}$ and $1.1 \times 10^6$ cm$^{-2}$ s$^{-1}$, respectively. 
A comparison of Jeans and non-thermal rates of escape of H$_2(v'=0,j')$ molecules, escaping in different rotational states $j'$ from martian dayside, is given in Figure \ref{fig4}.
To simplify the calculation we assumed the average solar conditions and neglected the latitude dependence of the production rate of hot O atoms. 
Jeans rate is about eight times greater than the non-thermal rate of the escaping H$_2$ for the lowest three rotational states, while for $j'>3$ the latter starts to dominate. The distinct character of the two RV distributions is a clear signature of different physical escape mechanisms. 

\begin{table}
\centering
\begin{tabular}{r | c  c  c}
\hline 
                                & H$_2$             & HD   & D$_2$   \\ \hline
   Jeans escape rate (s$^{-1}$) & $1.1 \times 10^6$ & 2.7  & $3.3 \times 10^{-6}$    \\
   Non-thermal escape rate (s$^{-1}$) & $1.9\times 10^5$  & 74   & 0.03 \\
\hline 
\end{tabular}
\caption{Total collisionally-induced escape rates of H$_2$, HD, and D$_2$ from the martian atmosphere.}
\label{table1}
\end{table}

While, in case of H$_2$, thermal rate is almost an order of magnitude higher than the collisionally-induced rate of escape, relative importance of the two processes changes for heavier isotopologues, namely HD and D$_2$ (Table \ref{table1}). A similar scaling of the collisional and thermal escape fractions can be expected for molecular escape from more massive astronomical objects.

\section{Conclusions}
We find that the collisionally-induced outflow of H$_2$ molecules and their heavier isotopes contributes to the evolution of the martian atmosphere. Namely, the escape rate of molecular hydrogen induced by collisions with hot oxygen from the martian atmosphere was calculated and found to be about six times smaller than the corresponding Jeans escape rate for the low solar activity. 
For heavier molecules, the collisional escape will dominate over thermal, as we have illustrated in case of HD and D$_2$ isotope molecules. In fact, the described process of molecular ejection induced by collisions may be one of the most important escape mechanisms of HD and D$_2$ from Mars. Consequently, the calculated escape fluxes of H$_2$ isotopologues could be important in analyses of the H/D ratio on Mars and evolution of water in martian history.

The described mechanism of molecular escape, where collisions provides sufficient translational energy to exceed the escape threshold and simultaneously excite internal molecular degrees of freedom, is rather general. It could be used to evaluate non-thermal escape fluxes and RV distributions of heavier molecules, such as CO, N$_2$, or CH$_4$, from Mars, Solar system bodies, and exoplanets. 
For the escape flux induced by O atoms produced in DR the upper limit on the mass of the escaping molecule is about $30 \; u$ on Mars.
Similarly, we estimate that the non-thermal escape of H$_2$ from a planetary atmosphere is possible for planetary masses up to about 3.4 Earth masses. A number of solar system bodies as well as the lightest currently confirmed exoplanets belong in that mass range. 
These limits do not include other non-thermal sources of hot atoms.

The escaping H$_2$ molecules exhibit a characteristic internal energy distribution, with a significant fraction of populated higher rotational states. Since H$_2$ molecules do not have a permanent electric dipole moment, they decay to the ground state mainly via collisions with the background gases present in an extended planetary corona. This is true for all escaping molecules that do not have ther permanent dipole moment. It could be possible to indirectly detect the presence of H$_2$ or other rotationally excited light molecules in the extended martian corona from a careful analysis of the collision rates and abundancies of the excited coronal species.
Finally, a significant amount of rovibrationally excited H$_2$ molecules remain in the martian atmosphere after colliding with hot O atoms. The cross sections and energy transfer parameters presented in this study can be used to determine non-thermal translational and rovibrational distributions of hot H$_2$ gas in the upper atmosphere of Mars.


%
%
%
%
%
%
%

\begin{acknowledgments}
We are grateful to D. Wang, A. Kuppermann, and J. Brand\~ao for providing Fortran subroutines for constructing potential energy surfaces, and to N. Lewkow for reading and suggestions. M.G. and V.K. were supported by NASA grants NNX09AF13G and NNX10AB88G.
\end{acknowledgments}

%
%
%
%
%
%
%
%
%

\bibliographystyle{agu08}








%

%
%

\end{article}

\end{document}